\newif\ifpdf
\ifx\pdfoutput\undefined
  \pdffalse
\else
  \pdfoutput=1
  \pdftrue
\fi

\ifpdf
  \documentclass[bulm,pdftex]{academic}
\else
  \documentclass[bulm]{academic}
\fi

\usepackage[tbtags]{amsmath}
\usepackage{amsfonts}
\usepackage{amssymb}

\renewcommand{\vec}[1]{\mbox{\boldmath $#1$}}
\newcommand{\mat}[1]{{\mbox{\bfseries \rmfamily#1}}}
\newcommand{\Idmat}{\vec{1}}
\newcommand{\diag}{{\rm diag}}

\newcommand{\fave}{\langle f \rangle}
\newcommand{\nuave}{\langle \nu \rangle}
\newcommand{\dave}[1]{\langle d^{(#1)} \rangle}
\newcommand{\mured}{\tilde\mu}

\begin{document}
\shortauthor{C. O. Wilke}
\shorttitle{Adaptive evolution on neutral networks}

\title{Adaptive evolution on neutral networks}

\author[1]{Claus O. Wilke%\footnote{Tel.: +1 626 395 2338, Fax.: +1 626 564
%    9651.\\Electronic submission is planned for the final version of the
%    manuscript.}
}
\address[1]{Digital Life Lab, Mail-Code 136-93\\
California Institute of Technology\\
Pasadena, CA 91125\\
wilke@caltech.edu}

\maketitle

\begin{abstract}
  We study the evolution of large but finite asexual populations evolving in
  fitness landscapes in which all mutations are either neutral or strongly
  deleterious. We demonstrate that despite the absence of higher fitness
  genotypes, adaptation takes place as regions with more advantageous
  distributions of neutral genotypes are discovered. Since these discoveries
  are typically rare events, the population dynamics can be subdivided into
  separate epochs, with rapid transitions between them. Within one epoch, the
  average fitness in the population is approximately constant. The transitions
  between epochs, however, are generally accompanied by a significant increase
  in the average fitness. We verify our theoretical considerations with two
  analytically tractable bitstring models.
\end{abstract}

\section{Introduction}

Sudden bursts of adaptive activity which punctuate long periods of stagnancy
seem to be a common observation in evolving systems. Such \emph{epochal
  evolution} \citep{vanNimwegenetal97}, has been found
in the fossil record \citep{EldredgeGould72,GouldEldredge77}, in evolving
bacteria \citep{LenskiTravisano94,Elenaetal96}, the evolution of tRNA
structures \citep{Huynenetal96,FontanaSchuster98} or artificial systems such as
digital organisms \citep{Adami95} and evolutionary
optimization \citep{VoseLiepins91,vanNimwegenetal97}. The most complete
theoretical analysis of epochal evolution has probably been presented in a
series of papers by van Nimwegen and
coworkers \citep{vanNimwegenetal97,vanNimwegenetal99a,%
vanNimwegenCrutchfield2000,vanNimwegenCrutchfield2000b}. The general picture
is as follows. A population can easily climb to the nearest local optimum, but
escape from there only with difficulty. Once a local optimum has been reached,
the population is trapped and experiences a metastable equilibrium. With a
relatively low probability, the population can discover a portal genotype
leading to the next local optimum, i.e., a genotype with a higher fitness than
what is currently present in the population. Once a portal is discovered, the
population moves quickly away from its current peak, and towards the new peak.
There, it settles down again in equilibrium, until the next portal is
discovered.

The above description focuses on the dynamics between local optima, but not on
the dynamics at one local optimum. However, in the presence of neutrality,
i.e., when a number of genotypes share the same identical replication rate, the
dynamics around one local optimum can be quite intriguing. A population does
not drift over a set of neutral genotypes, like a set of random walkers, but
has a quite different, even to some extent deterministic dynamics. In a
completely flat fitness landscape, for example, a population does not assume a
Gaussian distribution with a variance that increases over time, as would be
expected from a simple diffusion process. Rather, the population stays
clustered together and the cluster moves about as a whole
\citep{DerridaPeliti91}. If neutral and deleterious genotypes are mixed, i.e.,
if each genotype has as direct neighbors in genotype space both
neutral and deleterious genotypes, then the population moves to the regions of
the most connected neutral genotypes, as long as the mutation rate is small
\citep{vanNimwegenetal99b}.

In this paper, we study the dynamics of a finite but large population of
asexually replicating genetic sequences on a fitness landscape that contains
both neutral and strongly deleterious genotypes. We show that the neutral
genotypes naturally decompose into disjunct sets, and that an evolving
population can be trapped within such sets. Moreover, different sets yield
different reproductive success for the populations residing on them.  When a
population discovers a set of genotypes with higher reproductive success, the
population moves over to that set. For the average fitness of the population,
such transitions are reflected in a stepwise increase, exactly as it is
observed in standard scenarios of epochal evolution. However, the increase
observed here is not due to the discovery of faster replicating genotypes, but
solely to the discovery of genotypes with increased robustness against
mutations. Our results are valid for arbitrary mutation rates, and they
generalize the previous findings of \citet{vanNimwegenetal99b}.

\section{Theory}

We assume that there exist two classes of genotypes, those with a
relatively high replication rate $\sigma$, and the ones with a much lower
replication rate. For reasons of simplicity, we assume the latter to have
replication rate 0. This assumption is quite common in the literature
\citep{Gavrilets97,Gavrilets99,vanNimwegenetal99b}. Our analysis is similar to
the standard treatment of the quasispecies model
(see for example \citet{SchusterSwetina88,Eigenetal88,%
  Eigenetal89,Wilkeetal2001}) and to the work of
\citet{vanNimwegenetal99a,vanNimwegenetal99b}. Let the vector $\vec a$ define
the set of neutral genotypes, i.e., the genotypes with replication rate
$\sigma$:
\begin{equation}
  a_i = \left\{ \begin{array}{c@{\quad}l}
      1 & \mbox{if $i$ has replication rate $\sigma$,} \\
      0 & \mbox{else,}
    \end{array}\right.
\end{equation}
where $i$ runs over all possible genotypes. We assume a discrete time model,
and write the average population fitness as $\fave$. In
equilibrium, we have \citep{vanNimwegenetal99a}
\begin{equation}\label{quasisp-equil}
  \vec x = \frac{\sigma}{\fave} \mat Q \mat A \vec x\,,
\end{equation}
where $\vec x$ is the vector of concentrations, the diagonal matrix $\mat
A=\diag(a_0, a_1, \dots)$ contains the set of neutral genotypes, and the
matrix $\mat Q$ defines the mutation probabilities between different
genotypes, i.e., genotype $j$ mutates into genotype $i$ with probability
$Q_{ij}$.

In the following, we assume that the genotypes can be represented as sequences
of length $l$ over an alphabet of $A$ different symbols. Moreover, we assume a
uniform copy fidelity $q$ per symbol. That means, the $l$ symbols in a
sequence mutate independently from each other, and the substitution
probability in one generation is $1-q$ for each symbol. With this assumption,
the mutation matrix can be written as \citep{SwetinaSchuster82}
\begin{equation}
  Q_{ij} = q^l\left(\frac{1-q}{q(A-1)}\right)^{d(i,j)}\,,
\end{equation}
where $d(i,j)$ is the Hamming distance between sequences $i$ and $j$. It is
useful to introduce the reduced mutation rate $\mured$,
\begin{equation}
  \mured = \frac{1-q}{q(A-1)}\,,
\end{equation}
which allows us to write $\mat Q$ as a sum of matrices,
\begin{equation}\label{q-expans}
  \mat Q = q^l \sum_{k=0}^l \mured^k \mat D^{(k)}\,.
\end{equation}
The matrices $\mat D^{(k)}$ define the connection graphs at Hamming distance
$k$ in sequence space, i.e.,
\begin{equation}
  D_{ij}^{(k)} = \left\{ \begin{array}{c@{\quad}l}
      1 & \mbox{if $d(i,j)=k$,} \\
      0 & \mbox{else.}
    \end{array}\right.
\end{equation}
We insert \eqref{q-expans} into \eqref{quasisp-equil}, and obtain
\begin{equation}\label{equil-expans}
  \vec x = \frac{\sigma q^l}{\fave} \sum_{k=0}^l \mured^k
            \mat D^{(k)} \mat A \vec x\,.
\end{equation}

It is useful to introduce the matrices
\begin{equation}\label{conn-matrices}
  \mat G^{(k)} = \mat A \mat D^{(k)} \mat A\,.
\end{equation}
These matrices define the connection graphs at Hamming distance $k$ for the
neutral genotypes.
We now disregard all non-neutral sequences, and introduce
the concentration vector $\vec p$, which holds the concentrations of all
neutral sequences. The total number of neutral sequences is then
$P=\sum_i p_i$. Moreover, we assume all columns and rows corresponding to
non-neutral sequences to be deleted from the matrices $\mat G^{(k)}$.
From equation \eqref{equil-expans}, we obtain the eigenvalue equation
\begin{equation}\label{eigenvalue-1}
 \left(\frac{\fave}{\sigma q^l} -1\right) \vec p =
  \left(\sum_{k=1}^l \mured^k \mat G^{(k)}\right) \vec p\,.
\end{equation}
Consequently, the equilibrium state of the population is fully determined by
the matrix
\begin{align}\label{matrix-G}
  \mat G &= \sum_{k=1}^l \mured^{k-1} \mat G^{(k)}\notag \\
  &= \mat G^{(1)} + \mured \mat G^{(2)} + \mured^2 \mat G^{(3)} \dots
\end{align}
In the following, we will call matrices such as $\mat G$
\emph{generalized connection matrices}. The difference to the normal
connection matrices $\mat G^{(k)}$ as defined in equation~\eqref{conn-matrices}
is that generalized connection matrices may contain powers of $\mured$, i.e.,
they define a connection graph with weighted edges, whereas for the graphs
defined by $\mat G^{(k)}$, all edges have the same weight.

We can obtain further insight from relating the average population fitness
$\fave$ to the average fraction of neutral offspring, $\nuave$. Under the
assumption that the non-neutral sequences have a vanishing replication rate
(this assumption is equivalent to neglecting back mutations), we have
\citep{vanNimwegenetal99b}
\begin{equation}\label{nuave-and-fave}
  \nuave = \frac{\fave}{\sigma}\,.
\end{equation}
The probability $\nu_i$ for a single sequence $i$ to have a neutral genotype
as offspring is given by
\begin{align}\label{nu_i}
  \nu_i &= \sum_j a_j Q_{ji}\notag\\
    &= q^l \sum_{k=0}^l \mured^k\sum_j a_j  D_{ji}^{(k)} \notag\\
    &= q^l\left( 1+\sum_{k=1}^l \mured^k d_i^{(k)}\right)\,,
\end{align}
where $d_i^{(k)}$ gives the number of neutral neighbors at Hamming distance $k$
of sequence $i$. When we take the average over all viable sequences in the
population on both sides of equation \eqref{nu_i}, we arrive at
\begin{equation}\label{nu_ave}
  \nuave = q^l\left( 1+ \sum_{k=1}^l \mured^k \dave{k} \right)\,.
\end{equation}
The quantities $\dave{k}$ give the average number of neutral neighbors at
Hamming distance $k$ in the population. They can be expressed as
\begin{equation}
  \dave{k} = \frac{1}{P}\sum_i p_i\sum_j G_{ij}^{(k)}\,.
\end{equation}

With equation \eqref{nuave-and-fave}, we can rewrite equation \eqref{nu_ave} as
\begin{equation}\label{sum-neutralities}
  \sum_{k=1}^l \mured^k \dave{k} = \frac{\fave}{\sigma q^l} - 1\,.
\end{equation}
We notice that the right-hand-side of \eqref{sum-neutralities} is identical to
the factor in front of $\vec p$ on the left-hand-side of
\eqref{eigenvalue-1}. Therefore, we arrive at
\begin{equation}\label{eigenvalue-2}
  \left(\sum_{k=1}^l \mured^{k-1} \dave{k}\right) \vec p =
  \left(\sum_{k=1}^l \mured^{k-1} \mat G^{(k)}\right) \vec p\,.
\end{equation}
In the limit of a small mutation rate, \eqref{eigenvalue-2} becomes
\begin{equation}\label{vanNimwegen-result-general}
  \dave{1} \vec p = \mat G^{(1)} \vec p\,.
\end{equation}
In that case, the equilibrium distribution $\vec p$ depends solely on the
one-mutant connections of the neutral sequences, and the population neutrality
$\dave{1}$ is given by the spectral radius of the first-order connection
matrix. Equation~\eqref{vanNimwegen-result-general} corresponds to the result
of \citet{vanNimwegenetal99b}. Thus, we find that
equation~\eqref{eigenvalue-2} is the generalization of that result to
arbitrary mutation rates. For larger mutation rates, the equilibrium
distribution is influenced by the higher order connection matrices, and the
population neutrality may deviate from the spectral radius of $\mat G^{(1)}$.

\subsection{Separate neutral networks}
\label{network-hopping}

Following the conventional nomenclature in the
literature \citep{ForstReidysWeber95}, we call a set of neutral sequences
that can be transversed by one-point mutations a \emph{neutral network}. The
full set of neutral sequences in genotype space will in general decompose
into several disjunct such neutral networks. Assume there are $n$ disjunct
neutral networks. By reordering sequences, we can arrange $\mat G$ in
block-matrix form
\begin{equation}
  \mat G = \begin{pmatrix} \mat G_1 & & & &\\
 & \mat G_2 &  &\mathcal{O}(\mured) &\\
 &  & \mat G_3 & &\\
 &\mathcal{O}(\mured)  & &\ddots & \\
 &&&&\mat G_n \end{pmatrix}\,,
\end{equation}
where the matrices $\mat G_i$ are the generalized connection matrices for the
different neutral networks, and all off-diagonal terms are at least of the
order of $\mured$. Let us further assume that the matrices $\mat G_i$ are
ordered with descending spectral radius, i.e., the spectral radius of $\mat
G_i$ is larger than the one of $\mat G_{i+1}$ for all $i$. For a \emph{finite}
population initially residing on network $n$, we can then expect the following
dynamics. If the mutation rate is small, the population will equilibrate
within network $n$.  The discovery of a sequence that is part of a neutral
network $i<n$ is very unlikely, due to the small off-diagonal terms. However,
eventually such a sequence will be discovered. Since a progeny of that
sequence has a smaller probability to fall off its neutral network ($\mat G_i$
has a larger spectral radius than $\mat G_n$), the sequences on network $i$
will have a higher reproductive success than the sequences on network $n$, and
the population will move over to the newly discovered network. There, the
population will equilibrate, until the next higher-connected network is
discovered.  If the mutation rate is large, on the other hand, the
off-diagonal terms cannot be considered small anymore. In that case, the
discovery of higher connected regions is much more likely, and a population
will move straight away into the more densely connected regions of the
genotype space. In Sec.~\ref{sec:neutral-staircase}, we present examples for
both of these behaviors.

The above considerations show that the decomposition of the full set of
neutral sequences into neutral networks according to the definition given at
the beginning of this subsection is somewhat arbitrary. Depending on the
mutation rate, it might be justified, for example, to disregard contributions
of order $\mathcal{O}(\mured^2)$ to the matrix $\mat G$, but not the ones
of order $\mathcal{O}(\mured)$. In that case, it would be more natural to
group the sequences into sets that can be transversed by a combination of one-
or two-point mutations. In general, the neutral sequences should therefore be
subdivided into sets such that two arbitrary sequences of two disjunct sets
are at least a Hamming distance $k$ apart, where $k$ represents the
smallest number of simultaneous mutations that can be considered rare at the
respective mutation rate. For the remainder of this paper, we will understand
the term \emph{neutral network} in this more general sense.

\section{A simple exactly solvable landscape}

In this section, we study a simple example landscape for which the matrix
$\mat G$ can be diagonalized exactly, to all orders in $\mured$. We consider
binary sequences of length $l=2n$, and break them down into $n$ pairs of bits.
For each pair, we assume that there are three states (00, 01, 10) which are
neutral, and one state (11) which is lethal. Therefore, a sequence which
contains at least one pair for which both bits are set to 1 has a fitness of
0, and all other sequences have a fitness of $\sigma$. For simplicity, we set
$\sigma=1$. We will refer to this landscape as the \emph{Neutral Bitpairs}
landscape.

For a single pair, the matrix $\mat G_1$ reads (in this section, the subscript
$i$ in $\mat G_i$ indicates the number of pairs we are considering)
\begin{equation}
  \mat G_1 = \begin{pmatrix} 0 & 1 & 1 \\
    1 & 0 & \mured \\
    1 & \mured & 0 \end{pmatrix}\,,
\end{equation}
and its largest eigenvalue is $\lambda_1=(\mured +\sqrt{8 +\mured^2})/2$. For
a sequence with $2$ pairs, the corresponding matrix $\mat G_2$ can be written
as a tensor product (see \citet{Rumschitzki87,DressRumschitzki88,Wilke99}),
\begin{equation}
  \mat G_2 = \mured^{-1}\left[ (\Idmat + \mured\mat G_1 ) \otimes  (\Idmat +
    \mured\mat G_1 )
    - \Idmat\right]\,.
\end{equation}
The symbol $\Idmat$ stands for the identity matrix in the appropriate matrix
space. In general, for a sequence consisting of $n$ pairs, we can define the
matrix $\mat G_n$ recursively,
\begin{equation}
  \mat G_n = \mured^{-1}\left[ (\Idmat + \mured\mat G_1 ) \otimes  (\Idmat + \mured\mat G_{n-1} )
    - \Idmat\right]\,.
\end{equation}
As a consequence, the largest eigenvalue of $\mat G_n$ reads
\begin{align}
  \lambda_n &= \mured^{-1} [(1 + \mured \lambda_1)^n -1 ]\,,\notag\\
  &= \mured^{-1} [(1 + \frac{\mured^2}{2} + \frac{\mured}{2} \sqrt{8+\mured^2})^n-1]\,.
\end{align}
The average population fitness in this landscape follows from equation
\eqref{eigenvalue-1}. We obtain
\begin{equation}\label{ave-pop-fitness-simple-landscape}
  \fave = q^{2n}(1 + \frac{\mured^2}{2} + \frac{\mured}{2} \sqrt{8+\mured^2})^n\,.
\end{equation}
For small mutation rates, we can write the average fitness as
\begin{equation}\label{ave-fitness-approx}
  \fave = q^{2n}(1 + \sqrt{2}n\mured)\,,
\end{equation}
and the average population neutrality becomes
\begin{equation}\label{ave-pop-neutr}
  \dave{1} = \sqrt{2} n\,.
\end{equation}

\begin{figure}
\centerline{
\includegraphics[width=\columnwidth]{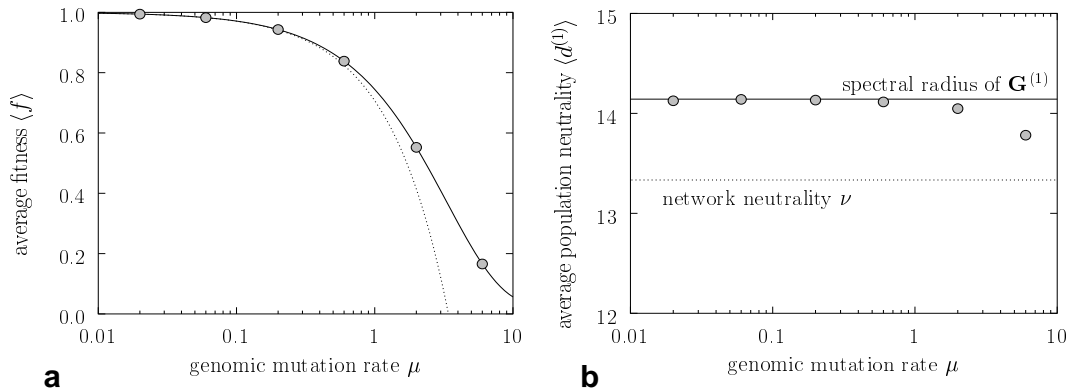}
}

\caption{\label{fig:neutr-bit-pairs} a: Average fitness as a function of
  the mutation rate in a Neutral Bitpairs landscape with $n=10$. The solid
  line represents equation~\eqref{ave-pop-fitness-simple-landscape}, the
  dotted line is the approximation equation~\eqref{vanNimwegen-result} with
  $\dave{1}$ given by equation \eqref{ave-pop-neutr}, and the points stem from
  simulations with a population of size $N=10000$. b: Average population
  neutrality $\dave{1}$ as a function of the mutation rate for the simulations
  from graph a. The solid line indicates the spectral radius of the first-order
  connection matrix $\mat G^{(1)}$, and the dotted line gives the network
  neutrality $\nu=\frac{2}{3}l$. }
\end{figure}

Let us now compare the full solution for the average population fitness to the
approximation given by \citet{vanNimwegenetal99b},
\begin{equation}\label{vanNimwegen-result}
  \fave = 1-\mu\left(1-\frac{\dave{1}}{l(A-1)}\right)\,,
\end{equation}
where $\mu$ is the genomic mutation rate,
\begin{equation}
  \mu = l(1-q)\,.
\end{equation}
Equation~\eqref{vanNimwegen-result} follows from
equation~\eqref{ave-fitness-approx} if we disregard all terms of the order
$\mathcal{O}\big((1-q)^2\big)$ or higher.  Figure~\ref{fig:neutr-bit-pairs}a
shows the exact solution for the average fitness $\fave$
equation~\eqref{ave-pop-fitness-simple-landscape} and the approximation
equation~\eqref{vanNimwegen-result} as a function of the genomic mutation rate
$\mu$. For comparison, we have also displayed results from numerical
simulations with a genetic algorithm. As was to be expected, the approximation
works well for $\mu\ll1$, but breaks down for higher mutation rates. For $\mu
\gtrsim 1$, the approximation significantly underestimates the average
population fitness. Interestingly, this goes along with a decrease in the
population neutrality (Fig.~\ref{fig:neutr-bit-pairs}b). What happens is the
following. In general, a population moves to that region of the genotype space
where the probability of neutral offspring is maximized for the given mutation
rate.  Clearly, when the mutation rate is low, only the immediate neighbors
influence that probability. If, however, the mutation rate is high, such that
offspring with two or even more mutations become common, the immediate
neighbors lose their importance. In the extreme case of a per symbol copy
fidelity of $q=0.5$, the probability of giving birth to a viable sequence
becomes identical for all sequences. Hence, in that extreme we can expect the
population neutrality to coincide with the network neutrality $\nu$ (the
network neutrality is the average number of neutral neighbors of a viable
sequence), which amounts to $\nu=\frac{2}{3}l$ in our case. At the same time,
most of the average population fitness does not stem from the one-mutants
anymore, but from mutants further away, which explains why the approximation
equation \eqref{vanNimwegen-result} underestimates the true average fitness.

\section{The Neutral Staircase landscape}
\label{sec:neutral-staircase}

\begin{figure}
\centerline{
\includegraphics[width=.9\columnwidth]{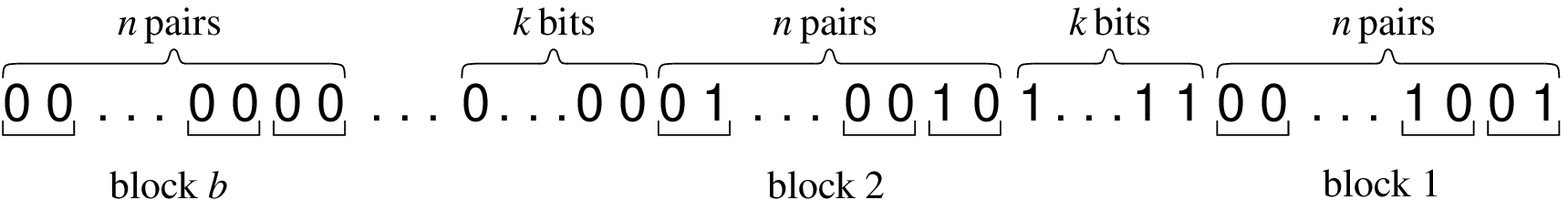}
}

\caption{\label{neutral-staircase}A valid string in the Neutral Staircase
  landscape. None of the two active blocks (block 1 and 2) contains a pair 11,
  all $k$ bits between blocks 1 and 2 are set to 1, and all bits to the left
  of block 2 are set to 0.  }
\end{figure}

The fitness landscape that we have studied in the previous section contains
only a single large neutral network, and hence epochal dynamics as predicted
in Sec.~\ref{network-hopping} cannot be observed in that landscape. However,
with a small modification, we can create a landscape that possesses the
required properties. We subdivide a binary sequence into $b$ blocks of length
$2n$, with a set of $k$ bits in between each block
(Fig.~\ref{neutral-staircase}).  The total length of the sequence is thus
$l=2bn+k(b-1)$. The blocks can be \emph{active} or \emph{inactive}. An active
block has properties similar to the sequences of the previous section. The
block is subdivided into $n$ pairs. If any of those $n$ pairs are in the
configuration 11, the fitness of the whole sequence is zero. Otherwise, the
fitness is not affected by that block.  The rightmost block is always active.
A block further to the left becomes active if the block immediately to its
right is active, and all $k$ bits between the two blocks are set to 1.
Finally, any bit to the left of the leftmost active block that is set to 1
results in fitness zero for the whole sequence. We call this landscape the
\emph{Neutral Staircase} landscape, in analogy to the \emph{Royal Staircase}
introduced by \citet{vanNimwegenCrutchfield2000b}. Note that the
Neutral Staircase differs from the Royal Staircase in an
important aspect: all sequences have either fitness 0 or fitness 1. No higher
fitness genotypes can be discovered, and the population's dynamics is
determined only by the topology of the neutral sequences.

The analytical treatment of the Neutral Staircase landscape is
straightforward for $k>1$. The neutral sequences decompose into $b$ neutral
networks, one for each possible number of blocks that can be active. Within
one neutral network, the average population fitness is readily available from
\eqref{ave-pop-fitness-simple-landscape}. Namely, if $i$ counts the number of
active blocks, we have
\begin{equation}\label{f-average-level-i}
    \langle f_i\rangle = q^{l}(1 + \frac{\mured^2}{2} + \frac{\mured}{2} \sqrt{8+\mured^2})^{ni}\,.
\end{equation}
Likewise, it is simple to derive the probability with which a new block is
discovered. First, we note that the probability with which an offspring
sequence remains on the neutral network $i$ is given by $\langle \nu_i\rangle
= \langle f_i\rangle$, according to \eqref{nuave-and-fave}. Hence, a single
offspring sequence ends up on the next neutral network with probability
$\langle f_i\rangle \mured^k q^{-2n}(1-q^2\mured^2)^n$ (we need $k$ extra
mutations to set the $k$ bits left of the leftmost active block to one, but we
are allowed some miscopies in the newly activated block). In a finite
population of $N$ sequences, there are on average $N\langle f_i\rangle$
sequences that can give rise to offspring. A sequence that belongs to the
network $i+1$ is therefore created in one generation with probability
\begin{equation}\label{crea-prob}
  P_{{\rm crea},i} = 1-[1-\langle f_i\rangle \mured^k
           q^{-2n}(1-q^2\mured^2)^n]^{N\langle f_i\rangle}\,.
\end{equation}
Next, we are interested in the probability of fixation of that newly
discovered sequence, $\pi_i$. For large populations, $\pi_i$ can be
approximated by $2s$ \citep{Haldane27,Kimura64}, where $s$ is the selective
advantage of the newly discovered network. In our case, we obtain thus
\begin{equation}
  \pi_i = 2\langle f_{i+1}\rangle/\langle f_i\rangle -1\,.
\end{equation}
The probability of a transition from network $i$ to network $i+1$,
$P(i\rightarrow i+1)$, is given by $P(i\rightarrow i+1) = P_{{\rm crea},
  i}\pi_i$, and hence the average epoch length $\tau_i$ follows as
\begin{equation}\label{tau-level-i}
  \tau_i = \frac{1}{P(i\rightarrow i+1)} = \frac{1}{P_{{\rm crea},i}\pi_i}\,.
\end{equation}
The time to convergence $\tau_{\rm conv}$, i.e., the time until the optimum
sequence distribution has been found, is given by
\begin{equation}\label{tau-conv}
  \tau_{\rm conv} = \sum_{k=1}^{b-1} \tau_k\,.
\end{equation}

\begin{figure}
\centerline{
\includegraphics[width=\columnwidth]{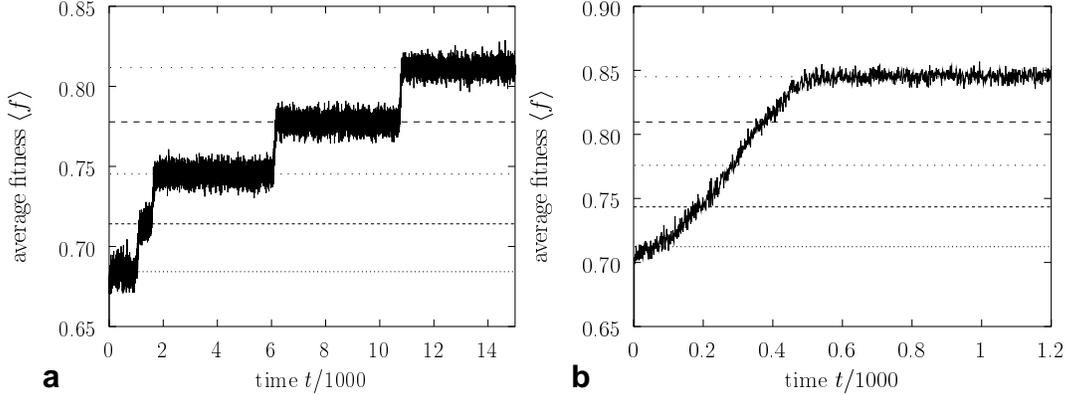}
}

\caption{\label{epochs}Two example runs on a Neutral Staircase landscape. If
  the entropic barrier between the neutral networks is large (a), the
  population moves from one network to the next in discrete jumps. If the
  barrier is small (b), the population moves straight away to the network that
  yields the highest average population fitness. We used $n=3$, $b=5$,
  $q=0.99$, and a population size of $N=10000$. In graph a, we used $k=3$, and
  in graph b, we used $k=2$.  }

\end{figure}

\begin{figure}
\centerline{
\includegraphics[width=9cm]{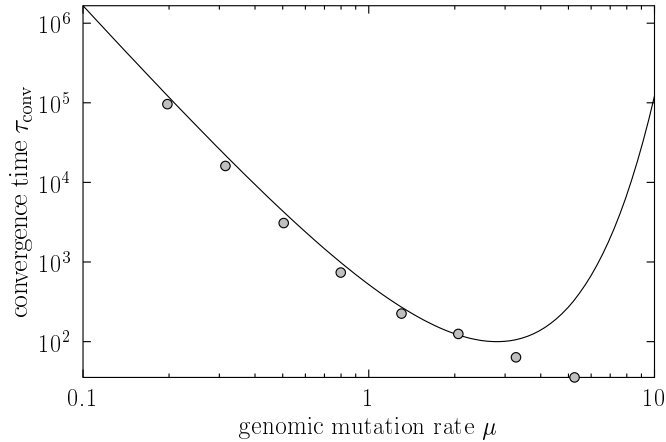}
}

\caption{\label{fig:conv-time}Convergence time $\tau_{\rm conv}$ in the
  Neutral Staircase landscape. The solid line is the analytical expression
  \eqref{tau-conv}, and the points are derived from 10 independent simulations
  each. The standard error of the measured convergence times is of the size of
  the symbols. We used $n=3$, $b=5$, $k=3$, and a population size of $N=10000$.
}

\end{figure}

Equations \eqref{f-average-level-i}, \eqref{tau-level-i}, and~\eqref{tau-conv}
capture most of the immediately observable quantities in an evolving
population in the Neutral Staircase landscape. We have tested their
applicability with numerical simulations of a genetic algorithm, and have
found good agreement. In Fig.~\ref{epochs}, we compare the average fitness in
the different epochs, $\langle f_i \rangle$, to the simulation results.
Figure~\ref{epochs}a shows that the population fitness fluctuates
around the predicted value during the metastable equilibrium, and that
transitions between two equilibria happen fast. In Fig.~\ref{epochs}b, we have
made the entropic barrier between two networks smaller (we have reduced $k$).
Then, all neutral sequences actually form only a single large neutral network,
and the population transitions immediately to the highest possible level of
the average fitness.

In Fig.~\ref{fig:conv-time}, we compare the time to convergence $\tau_{\rm
  conv}$ with simulation results. We find that for $\mu\lesssim 3$, our
analytical prediction agrees well with the simulations, whereas for larger
$\mu$, the prediction fails to capture the observed behavior. The origin of
this failure is the following. We have derived equation~\eqref{tau-conv} under
the assumption that it is hard to discover the next higher network. Now, when
$\mu$ becomes of the order of $k$ (we have $k=3$ in Fig.~\ref{fig:conv-time}),
mutations that flip all $k$ bits at once to activate the next block become
likely, and hence this assumption fails. As in the case of Fig.~\ref{epochs}b,
the neutral sequences then form a single gigantic network, and an expression
derived under the assumption of disjunct sub-networks must break down.

\section{Discussion}
The analysis presented in the preceding sections has shown that the topology
of the neutral sequences in genotype space has an important influence on the
dynamics of a population. It is therefore not justified to regard the
evolution of a population of neutral genotypes as a simple diffusion
process, unless either the population size or the mutation rate are very small
(when the product of the population size, $N$ and the genomic mutation rate,
$\mu$, is much smaller than one, the population as a whole performs
essentially a random walk on the neutral network \citep{vanNimwegenetal99b}.)
We have demonstrated that evolution on neutral networks can lead to the same
kind of epochal dynamics that previously was thought to be caused solely by
the discovery of genotypes of higher fitness. Of course, our results do not
imply that sudden transitions in a population are never caused by such
discoveries. Normally, a transition that we observe in a population evolving
in an unknown fitness landscape will be due to the discovery of a faster
replicating genotype. However, sometimes we may observe a transition to a
higher average fitness without an increase (or even with a decrease, see
below) in the fitness of the dominant or fastest replicating genotype. Such a
transition is then due to the effects described in this paper.

In most cases, selection acts first and foremost on replication rates.
However, if all viable genotypes are identical in terms of their replication
rate, as is the case in the fitness landscapes we have studied in this paper,
the next important quantity selection acts upon is the probability with which
genotypes have viable offspring. This probability is more a property of a set
of genotypes than of a single genotype, because a genotype that itself has
high reproductive success, but produces mainly offspring with poor
reproductive success, will ultimately have only a small number of progeny.
Therefore, the selective pressure we have described here acts solely on clouds
of mutants, in an extreme form of quasispecies-like selection
\citep{EigenSchuster79,Nowak92}.

Throughout this paper, we have assumed that all sequences have the same
replication rate $\sigma$. It is natural to ask what happens when sequences
with different replication rates are present. The simplest such situation
occurs when all sequences within a single neutral network $i$ have the same
replication rate $\sigma_i$. In that case, we can take the analysis given in
Sec.~\ref{network-hopping} one step further. The equilibrium fitness $\langle
f_i\rangle$ within a network $i$ is then given by
\begin{equation}
  \langle f_i\rangle = \sigma_i q^l(1+\mured \rho_i)\,,
\end{equation}
where $\rho_i$ is the spectral radius of the connection matrix of the network,
$\mat G_i$. A population will of course try to move to the particular network
that gives the maximum equilibrium fitness. For a small mutation rate
$\mured$, it is clear that the network with the largest $\sigma_i$ yields the
maximum equilibrium fitness. If, however, the mutation rate is large, such
that $\mured \rho_i$ becomes of the order of unity or even exceeds that value,
then depending on the distribution of the $\rho_i$'s and the $\sigma_i$'s over
the different networks, a population may actually transition from a network
$j$ with a higher replication rate $\sigma_j$ to a network $i$ with a smaller
replication rate $\sigma_i$, while at the same time increasing its average
fitness. A similar effect was already predicted by \cite{SchusterSwetina88},
although they investigated peaks with different support from slightly
deleterious mutations, rather than neutral networks with different connection
densities.  In recent work, \citet{Ofriaetal2000} have observed this
selection for neutrality empirically in digital organisms. In a large number
of evolution experiments with different mutation rates, they observed that the
number of neutral neighbors of the dominant genotype (a crude but practical
measure for the spectral radius of the connection matrix) would increase at
higher mutation rates, at the expense of the digital organisms' replication
rate. Similarly, \citet{Wilkeetal2000} showed that digital organisms with a
vastly inferior replication rate could outcompete seemingly superior digital
organisms at high mutation rates if the slower replicating organisms had a
higher robustness against mutations.

It is interesting to relate our results for the average population fitness to
the mutational load $L$, which is defined as \citep{Haldane37,Muller50,Crow70}
\begin{equation}
  L = 1-\frac{\fave}{\sigma}\,.
\end{equation}
From equation~\eqref{eigenvalue-1}, we obtain
\begin{equation}\label{mut-load-1}
  L = 1 -q^l(1+\mured \rho)\,,
\end{equation}
where $\rho$ is the spectral radius of the connection matrix $\mat G$. As was
already noted by \citet{vanNimwegenetal99b}, the load can deviate
significantly from Haldane's result $L=\mu$ \citep{Haldane37} if neutrality is
present [note that equation~\eqref{mut-load-1} becomes identical to Haldane's
result in the absence of neutrality ($\rho=0$) and in the limit of a small
mutation rate]. More importantly, as a generalization of Haldane's result, it
is often cited that the genetic load in an asexual population is independent
of the fitness landscape, and therefore also of epistasis, and that it is
equal to
\begin{equation}\label{mut-load-2}
  L = 1-e^{-\mu}.
\end{equation}
(see e.g.\ \citet{Kondrashov88,Charlesworth90}, derivation given by
\citet{KimuraMaruyama66,Crow70}). This result, however, holds only in the
absence of neutral mutants. If neutral mutants are present, then the topology
of the neutral genotypes in the genotype space is coupled to the mutational
load, by virtue of $\rho$ in equation~\eqref{mut-load-1}. Since the topology
is also coupled to the type of epistasis that we observe
\citep{WilkeAdami2000}, epistasis must have an influence on the mutational
load, even in asexual populations, as long as neutral mutations can occur.
However, equation~\eqref{mut-load-2} is the basis of the deterministic
mutational hypothesis of the evolution of sex \citep{Kondrashov88}: if the
mutational load is independent of epistasis for asexual populations, but
depends strongly on the sign of epistasis for sexual populations, then for
certain types of epistasis the mutational load of a sexual population may be
much smaller than that of an asexual population. Because of the differences
between equations~\eqref{mut-load-1} and~\eqref{mut-load-2}, the
deterministic mutational hypothesis could break down if neutrality were taken
into account. As a consequence, this hypothesis should
be reconsidered for the case of landscapes with neutrality.

\section{Conclusions}

We have shown that adaptive evolution can take place in the complete absence
of what is ordinarily understood as advantageous genotypes. Even if the
fitnesses of all viable genotypes are completely identical does selection
favor particular regions of genotype space over others. What makes the
difference is the density of neutral sequences. In a region of genotype space
with a higher density of neutral sequences, chances are higher that a mutated
offspring is neutral rather than deleterious. Therefore, neutral sequences in
such a region have a higher robustness against mutation, and hence a higher
reproductive success. This gives them sufficient selective advantage to
outcompete sequences from a less densely connected region. The transitions
between different such regions will often occur in sudden jumps, followed by
relatively long periods of stasis. Evolution on neutral networks alone can
thus lead to epochal dynamics observed in so many natural and artificial
evolving systems.

\section*{Acknowledgements}
We thank Chris Adami for carefully reading this manuscript. This work was
supported by the NSF under contract DEB-9981397.

\frenchspacing

\bibliography{paper.bib}
\bibliographystyle{plainnat}

\end{document}